\documentclass[12pt]{article}

\usepackage[letterpaper,hmargin=1in,vmargin=1in]{geometry}

\usepackage{graphicx,epstopdf,amsmath,amsfonts,amssymb}

\def\be{\begin{equation}}
\def\ee{\end{equation}}
\def\ba{\begin{eqnarray}}
\def\ea{\end{eqnarray}}
\def\ge{\mathrel{\raise.3ex\hbox{$>$\kern-.75em\lower1ex\hbox{$\sim$}}}}
\def\la{\mathrel{\raise.3ex\hbox{$<$\kern-.75em\lower1ex\hbox{$\sim$}}}}

\def\simgt{\mathrel{\raise.3ex\hbox{$>$\kern-.75em\lower1ex\hbox{$\sim$}}}}
\def\simlt{\mathrel{\raise.3ex\hbox{$<$\kern-.75em\lower1ex\hbox{$\sim$}}}}

\newcommand{\bi}[1]{\bibitem{#1}}
\newcommand{\fr}[2]{\frac{#1}{#2}}

\newcommand{\nc}{\newcommand}

\nc{\gone}{\bar g_{\pi NN}^{(1)}}
\nc{\gzero}{\bar g_{\pi NN}^{(0)}}
\nc{\al}{\alpha}
\nc{\ga}{\gamma}
\nc{\de}{\delta}
\nc{\ep}{\epsilon}
\nc{\ze}{\zeta}
\nc{\et}{\eta}
\nc{\ka}{\kappa}
\nc{\rh}{\rho}
\nc{\si}{\sigma}
\nc{\ta}{\tau}
\nc{\up}{\upsilon}
\nc{\ph}{\phi}
\nc{\ch}{\chi}
\nc{\ps}{\psi}
\nc{\om}{\omega}
\nc{\Ga}{\Gamma}
\nc{\De}{\Delta}
\nc{\La}{\Lambda}
\nc{\Si}{\Sigma}
\nc{\Up}{\Upsilon}
\nc{\Ph}{\Phi}
\nc{\Ps}{\Psi}
\nc{\Om}{\Omega}
\nc{\ptl}{\partial}
\nc{\del}{\nabla}
\nc{\ov}{\overline}
\nc{\newcaption}[1]{\centerline{\parbox{15cm}{\caption{#1}}}}
\nc{\us}{U(1)$_S$}

\def\beq{\begin{equation}}
\def\eeq{\end{equation}}
\def\bmat{\begin{displaymath}}
\def\emat{\end{displaymath}}
\def\bear{\begin{eqnarray}}
\def\eear{\end{eqnarray}}
\def\ba{\begin{eqnarray}}
\def\ea{\end{eqnarray}}
\def\bery{\begin{array}}
\def\ery{\end{array}}
\def\bit{\begin{itemize}}
\def\eit{\end{itemize}}
\def\ben{\begin{enumerate}}
\def\een{\end{enumerate}}
\def\btab{\begin{tabular}}
\def\etab{\end{tabular}}
\def\btbl{\begin{table}}
\def\etbl{\end{table}}
\def\bfig{\begin{figure}[htb]}
\def\efig{\end{figure}}
\def\bpic{\begin{picture}}
\def\epic{\end{picture}}

\def\ga{\mathrel{\raise.3ex\hbox{$>$\kern-.75em\lower1ex\hbox{$\sim$}}}}
\def\la{\mathrel{\raise.3ex\hbox{$<$\kern-.75em\lower1ex\hbox{$\sim$}}}}
\def\gappeq{\mathrel{\rlap {\raise.5ex\hbox{$>$}}
{\lower.5ex\hbox{$\sim$}}}}
\def\lappeq{\mathrel{\rlap{\raise.5ex\hbox{$<$}}
{\lower.5ex\hbox{$\sim$}}}}

\def\gyr{{\rm \, G\kern-0.125em yr}}
\def\mev{{\rm \, Me\kern-0.125em V}}
\def\gev{{\rm \, Ge\kern-0.125em V}}
\def\tev{{\rm \, Te\kern-0.125em V}}

\begin{document}

\begin{titlepage}

\setcounter{page}{1}

\vspace*{0.2in}

\begin{center}

\hspace*{-0.6cm}\parbox{17.5cm}{\Large \bf \begin{center}

Exploring Portals to a Hidden Sector \\
Through Fixed Targets

\end{center}}

\vspace*{0.5cm}
\normalsize

\vspace*{0.5cm}
\normalsize

{\bf Brian Batell$^{\,(a)}$, Maxim Pospelov$^{\,(a,b)}$, and Adam Ritz$^{\,(b)}$}

\smallskip
\medskip

$^{\,(a)}${\it Perimeter Institute for Theoretical Physics, Waterloo,
ON, N2J 2W9, Canada}

$^{\,(b)}${\it Department of Physics and Astronomy, University of Victoria, \\
     Victoria, BC, V8P 1A1 Canada}

\smallskip
\end{center}
\vskip0.2in

\centerline{\large\bf Abstract}

We discuss the sensitivity of neutrino experiments at the luminosity frontier 
to generic hidden sectors containing new (sub)-GeV
neutral states.  The weak interaction of these states with the Standard Model can be efficiently probed
through all of the allowed renormalizable `portals'  (in the Higgs, vector, and neutrino sectors) at fixed target
proton beam facilities, with complementary sensitivity to colliders. 
We concentrate on the kinetic-mixing vector portal, and show that certain regions of
the parameter space for a new \us\ gauge sector with long-lived sub-GeV mass states 
decaying to Standard Model leptons are already severely constrained 
by the datasets at LSND, MiniBooNE, and NuMI/MINOS.  Furthermore, 
scenarios in which portals allow access to stable neutral particles, such as MeV-scale dark matter, generally 
predict that the neutrino beam is accompanied by a `dark matter beam', observable through neutral-current-like 
interactions in the detector. As a consequence, we show that the LSND electron recoil 
event sample currently provides the most stringent direct constraint on MeV-scale dark matter models.

\vfil
\leftline{June 2009}
    
\end{titlepage}

\subsection*{1. Introduction}

The scale at which new physics is to be expected is not always straightforward to ascertain from empirical data.
The generic arguments for uncovering a Higgs (or Higgs-like) sector at the weak scale, in tandem with additional structure that alleviates the fine-tuning issues of the Standard Model (SM), leads to the pervading notion that all new physics should be found at
the high-energy frontier. In part, this results from strong existing constraints on new physics charged under the 
SM gauge group, as would generically be the case in the context of electroweak symmetry 
breaking. However, one of the most compelling empirical arguments for new physics is the need to explain dark matter (DM),
which suggests instead the presence of a hidden sector,  uncharged (at low energies) under the SM gauge group. The mass scale
of the states in a hidden sector, even if the dark matter candidate itself is a WIMP with a weak-scale mass, is far less
constrained. The notion of a hidden sector is also a generic feature of supersymmetric models of the weak scale, 
 due to the specifics of supersymmetry breaking and mediation. It is also notable that the presence of new physics in a hidden (uncharged)
 sector is quite compatible with the strong hints toward unification of the SM gauge groups at a high scale.

It is worthwhile recalling that, even in the SM, some of the matter fields are uncharged under one or more of the color and electroweak
 gauge groups, and thus the extension to include a further sector is not particularly exotic from a theoretical perspective, even beyond the empirical
motivations noted above. However, it is clear that direct sensitivity to  a neutral hidden sector will depend in detail on the couplings,
which may  be associated with higher mass states charged under the SM that have been integrated out, as well as
possible light neutral states present at low energies. Thus it is helpful to ask in which regimes we have the best experimental sensitivity. 
With respect to the mass and lifetime range for hidden sector states, the niche 
with masses in the range of a few MeV to a few GeV, with lifetimes of less than 1 second, is quite intriguing as many of the 
more severe astrophysical and cosmological constraints that apply to lighter states are weakened or eliminated, while those
from high-energy colliders are often inapplicable. Certain classes of hidden sectors may be strongly constrained by flavor
physics and other precision tests, but it is clear that many generic scenarios are viable. Recent hints from various astrophysical
anomalies, when interpreted in the context of dark matter, also point to this range which suggests that probing
hidden sectors with light (approximately) flavor-conserving MeV-to-GeV scale states weakly coupled to the SM model represents 
an intriguing possibility.

Given a hidden sector, which at low energies necessarily only contains states that are electrically neutral and presumably
neutral with respect to the full SM gauge group, we can parameterize the interactions as follows,
\be
\label{int}
{\cal L}_{\rm mediation} = \sum_{k,l,n}^{k+l = n+4} \fr{{\cal O}^{(k)}_{\rm  NP} {\cal O}^{(l)}_{\rm SM} }{\Lambda^n},
\ee
where ${\cal O}$ denotes SM and new physics (NP) operators of canonical dimensions $k$ and $l$, and 
$\Lambda$ is a cutoff scale presumably at a TeV or above. In general  ${\cal O}_{\rm NP}$ and ${\cal O}_{\rm SM}$ 
can transform via various Lorentz representations, but as discussed above we will consider the situation
where ${\cal O}^{(l)}_{\rm SM}$ is gauge invariant under the SM gauge group.
The case of marginal $n=0$ interactions is especially important and will be the focus of attention here.
The SM operators of lowest dimension are well known and are often called {\it portals} \cite{pw,portal}. These operators include:
\begin{eqnarray}
F^Y_{\mu\nu}~~~~~~~~~~~&&{\rm Vector~portal~(dim=2),} 
\nonumber
\\ H^\dagger H ~~~~~~~~~~~&&{\rm Higgs~portal~(dim=2),} 
\label{portals}
\\ 
L H ~~~~~~~~~~~&&{\rm Neutrino~portal~(dim =5/2)},
\nonumber 
\end{eqnarray}
where $F^Y_{\mu\nu}$, $ H$ and $L$ are the hypercharge field strength, and the Higgs and lepton doublets. 
The operators (\ref{portals}) allow a coupling of the SM to (SM neutral) new physics at the renormalizable level,
and thus do not presuppose anything about the mass scale for these new fields. Note that the
existence of neutrino oscillations can be viewed as new physics entering through the neutrino portal,
e.g. in the form of a coupling to right-handed neutrinos. It is then reasonable to ask whether other portals are 
realized in nature, and whether we have the ability to probe them experimentally.

Opportunities to probe hidden sectors experimentally arise in particular when states exist with a mass scale of a  
GeV or less and a lifetime longer than the mesons which decay due to weak interactions. Such light, long-lived hidden sector states can be searched for at
high-luminosity fixed target experiments where the large collision-number
statistics  can overcome the (assumed) weakness of the interaction, allowing these states to be produced in sufficient 
numbers.  A rough comparison between the collider and fixed target 
reach for new light states in neutral hidden sectors can be made as follows. Assuming the interaction between the two sectors is 
mediated by marginal or irrelevant operators of dimension $4+n$ with 
$n \geq 0$ as shown in Eq.~(\ref{int}), the production cross section will typically scale as 
\be
\sigma \sim \fr{\kappa^2}{E^2}\left(\fr{E}{\Lambda}\right)^{2n},
\label{sigmascale}
\ee
where $\kappa$ is a dimensionless coupling constant and $\Lambda$ is a large ({\em e.g.} TeV) scale
which determines the intrinsic weak-coupling `barrier' to probing the portal. 
While the characteristic integrated luminosity for  high-energy colliders is of order 
$10^{41}{\rm cm}^{-2}$, the analogue of integrated luminosity for a fixed  target of length 1m with $10^{21}$ protons on
target (POT) is $10^{21} \times 10^{24} {\rm cm}^{-3} \times  10^2 {\rm cm} \sim 10^{47}{\rm cm}^{-2}$.
With these numbers, one arrives at the following comparison between the production rates for neutral GeV-scale states 
at colliders and fixed targets:
\be
\fr{N_{\rm collider}}{N_{\rm target}} \sim 10^{-6} \times \left( \fr{E_c}{E_t}\right)^{2n-2} 
\sim 10^{-12 + 6n},
\label{counting}
\ee
where in the last relation we have assumed that the attainable collider energies are three orders of magnitude larger than 
$s^{1/2}$ for fixed target experiments, $E_{\rm c}/E_{\rm t} = E_{\rm c}/\sqrt{2m_pE_{\rm lab}} \sim 10^{3}$ for $E_{\rm c} =14$ TeV and 
$E_{\rm lab} =100$ GeV. For low $n=0,~1$ the production rate at fixed targets is clearly advantageous, and even at $n=2$ 
can be comparable to colliders. 
However, this does not directly imply that the fixed target signal would necessarily be larger, due to the
 geometric acceptance criteria of fixed target detectors. In contrast, if the collider signal consists
primarily of missing energy, detection may in turn be difficult
due to the size of the SM background. In essence, (\ref{counting}) demonstrates 
that for $n \leq 2$ fixed targets may very well be ahead of colliders in sensitivity to new GeV-scale hidden sector 
physics, although the precise comparison clearly depends on the specific model under consideration.

Our primary observation here is that the existing and ongoing development of experimental infrastructure
for probing neutrino physics through beams generated by high-intensity proton sources directed on fixed targets 
is also an ideal setting within which to probe all the SM portals, and couplings to a hidden sector discussed above. 
As we will describe in more detail below, past and existing experiments ranging in beam energy from LSND to MiniBooNE and MINOS, and those in development such as T2K and NOvA, have strong sensitivity to portal couplings.
Of course, the idea to search for exotic (un)stable states using fixed targets is far from new 
and in the past several experimental studies have been performed 
targeting different types of exotics \cite{PDG}. In particular, we note the existing
experimental searches for axions \cite{axions_exp},  heavy neutrino states \cite{neutrino_exp}, 
long-lived gluinos \cite{gluino_exp}, unstable light particles in general \cite{unstable}, and 
light particles with neutral-current-type interactions \cite{losecco}. 
Theoretical motivation for these searches often originated from several unrelated considerations, 
see {\em e.g.} \cite{WW,Shrock,Farrar,Dreiner,neutrino_th} for an incomplete list. 
We stress, however, that the majority of these searches were performed in the 80s and 90s with typical POT$\sim10^{18}$, which allows for significant improvements with modern facilities where the proton exposure can be 
above POT$\sim 10^{21}$. Moreover, leaving particular theoretical scenarios aside, the description of interactions in terms of SM portals  
allows the GeV-scale phenomenology of a hidden sector to be systematized. 

In recent years, the scope for particle physics interpretations
of dark matter,  observed through its gravitational impact on cosmology and astrophysics 
over many time and distance scales, has widened considerably. 
Among numerous dark matter scenarios, several interesting possibilities have been proposed
that deviate from the most commonly assumed candidate of an electroweak-scale WIMP 
coupled to the SM via higher-dimensional contact operators. 
Notably, models of MeV-scale dark matter coupled to the SM via a sub-GeV mediator 
\cite{BF,Fayet,PRV,HZ}
have been proposed in an attempt to explain the  511 keV emission observed
from the galactic center with annihilating dark matter \cite{Boehm}. Another  suggestion 
invoked WIMPs at the GeV scale \cite{light-chi} to explain the results of the DAMA experiment \cite{DAMA}. 
Most recently, it was suggested that (sub-)GeV scale mediators between the dark and visible 
sectors could enhance the annihilation cross section of electroweak scale WIMPs to  
leptons \cite{AFSW,PR}, in tantalizing agreement with the positron excess observed by 
PAMELA  \cite{Pamela}. While many astrophysical anomalies may find alternative explanations, 
the variety of new effects that may occur in 
dark matter phenomenology due to the presence of light mediators or dark matter particles 
has led to a renewed interest in their low and intermediate energy particle physics manifestations 
 \cite{Fayet,Drees,tests,BPR,slac,Reece}. However,
the sensitivity of these tests often decreases when the signal contains long-lived particles 
that escape the detector. In such cases, fixed target experiments with
detectors 10m--1km from the target, as in modern long-baseline
neutrino experiments, can provide complementary sensitivity.
A rather striking consequence of models with light dark matter is the production of a high intensity `dark matter beam', generated as dark matter particles are pair-produced as a result of the proton-target interactions and boosted along the proton beam direction.
The scattering of WIMPs of this type in the (near-)detector 
would generate an additional contribution to neutral-current type events (see, {\em e.g.} \cite{losecco}). This prediction implies that a
direct search for MeV-scale stable WIMP particles is possible at experiments at the luminosity frontier. 

This paper aims to demonstrate that the physics reach of  fixed target experiments, e.g.
in the conventional neutrino beam plus near-detector set up, surpasses the reach of 
previous experimental probes and covers a wider range of possibilities 
for new physics coupled via the vector and Higgs portals. We consider the 
kinetic-mixing vector portal in detail and estimate the number of decays of new GeV and sub-GeV-scale 
states in past and existing detectors such as LSND, MiniBooNE and NuMI/MINOS. 
In particular, we show that certain parts of the parameter space,
inaccessible to collider experiments, are already strongly disfavored or ruled out by 
fixed target experiments as a by-product of the neutrino physics analysis. Moreover, 
we demonstrate that  DM beam scattering inside the LSND detector places strong restrictions 
on MeV-scale dark matter models, superior to all pre-existing limits. 

The remainder of this paper is organized as follows. In the next section we provide some generic details 
on the coupling of GeV-scale neutral hidden sector states via the most important portals. We then proceed
to analyze in detail the 
signatures of long-lived vectors and Higgs$'$ bosons from the secluded \us\ sector in
Section 3. Section 4 contains the calculation of the neutral-current-like signal 
of an MeV-scale DM beam scattering inside the detector, while Section 5 provides a discussion 
of possible future improvements in fixed target probes of hidden sectors.

\subsection*{2. Standard Model portals and GeV-scale states}

Focussing on the portals in (\ref{portals}) allows a classification of the
interactions of new GeV-scale states with the SM. In this section, we
discuss the various cases and note regions of parameter space
leading to relatively long lifetimes, as appropriate to exploration
at fixed target facilities.

\subsubsection*{2.1 Vector portal} 

The simplest way  in which the SM may interact with a new abelian gauge sector
was pointed out long ago \cite{holdom},
\be
{\cal L}_{\rm int}=\frac{\kappa}{2}\, V_{\mu\nu} F^{\mu\nu},
\ee
where $F^{\mu\nu}$ and $V_{\mu\nu}$ are the SM electromagnetic and secluded U(1)$_S$ field strengths. At higher
energies, $F^{\mu\nu}$ should be replaced with the hypercharge field strength, but we
will ignore the coupling to the $Z$ in the energy range considered here. 
After the spontaneous breaking of the secluded \us, the mixing parameter $\kappa$ appears in 
front of  the coupling of $V_\mu$ to the electromagnetic current,
\be
{\cal L}_{\rm int}=\kappa V_{\mu} J^{EM}_{\mu} + \frac{m_V^2}{v'} h' V_\mu^2+\cdots.
\ee
We have also retained the interaction of the secluded Higgs$'$ ($h'$) and $V$-bosons, which will be important for signatures involving $h'$. Although the appeal of this model is in its extreme 
simplicity, there are nevertheless a
considerable range of intricate phenomenological 
consequences (see, for example, \cite{tests,BPR}). Such models have recently been under
scrutiny in situations where the WIMP dark matter candidate is also 
charged under \us\ and the light vector leads to a Sommerfeld-enhanced annihilation
rate.

There are two distinct regimes that allow for a long-lived particle in the \us\ sector. 
The first corresponds to $\kappa \la 10^{-7}-10^{-6}$, when the sub-GeV vectors 
may have a decay length $c\tau$ in excess of several meters. In this case the secluded 
Higgs$'$ must be heavier than the vector, $m_{h'} > m_V$, allowing for a prompt Higgs$'$ decay. 
Another distinct possibility is when $\kappa \sim 10^{-4}-10^{-2}$ and $m_V > m_{h'}$. 
This is the case of a fast decaying vector, but a long-lived Higgs$'$ with a decay width of second 
order in the mixing parameter $\kappa$ \cite{tests,PRV}. Both possibilities are ideal candidates 
for a fixed target search, and will be analyzed in some detail in the next section.

\subsubsection*{2.2 Higgs portal} 

Interactions through the Higgs portal, namely couplings to $H^\dagger H$ \cite{pw,portal} are distinct in the sense that
the Higgs sector of the SM has not been observed, and thus its detailed features are unknown. Thus, the 
Higgs portal is simply one parameterization of an extended Higgs sector, and we can take the new state
to be a scalar singlet  $S$, that may couple to the Higgs portal  via dimension 3 and 4 operators,
\be
{\cal L}_{\rm int} = (H^\dagger H)(\lambda S^2+A S)=
 hv(\lambda S^2 + AS) +\cdots,
\ee 
where after the spontaneous breaking of the electroweak symmetry, $H^T = (0, (v+ h)/\sqrt{2})$,
and only the interaction terms with the Higgs boson are retained. When $A=0$ the model has an additional 
$Z_2$ symmetry so that the particles $S$ are stable and are a viable DM candidate \cite{singlet1,singlet2}.
The light mass scale for $S$ has implications for the natural size of $\lambda$. Typically, one 
would require $\lambda v^2 \la m_S^2$ in order to avoid finely-tuned cancellations. This is not necessarily 
the case for the two Higgs-doublet model with a singlet  at large $\tan\beta$, as the coupling of $S^2$ 
to $H_d^\dagger H_d$ does not have to be very small from naturalness arguments alone \cite{Bird}.
Keeping to the minimal model, and treating $\lambda$ and $A$ as free parameters, we can integrate out 
the physical Higgs field as we are interested in low energy phenomenology. The result can 
be written in the following form, 
\be
\label{eff_h}
{\cal L}_{\rm int} = {\cal O}_{\rm SM}^{(h)}\fr{ \lambda S^2 + AS}{m_h^2},
\ee
where ${\cal O}_{\rm SM}^{(h)} = \sum_f m_f \bar f f +...$ is the familiar ($q^2$-dependent) SM 
operator describing the interaction of an off-shell Higgs of momentum $q$ 
with the SM fermions, gluons and photons. It is important to keep in mind 
that unlike the vector case where flavor physics does not impose very restrictive constraints
on the couplings, the Higgs portal can be constrained via rare $B$ and $K$ decays 
\cite{BtoK,Bird,PRV}.

Long lifetimes for $S$ particles can arise either by symmetry or by kinematic accident.
In the case of an approximate $Z_2$ symmetry, $A\ll v$, the smallness of $A$ 
ensures the longevity of $S$. The lifetime of $S$ is given by the lifetime of the Standard Model
Higgs boson at $m_h = m_S$ rescaled by the small mixing angle $\ka'\equiv Av/m_h^2$:
\be
\Gamma_S = (\kappa')^2\Gamma_h(m_h=m_S).
\ee
In order to achieve a long lifetime (e.g. in excess of $10^{-6}$ s), one has to require a very small 
value for $A$ or to ensure that $m_S$ is below the di-muon threshold. In this case the 
decay to electrons or $\gamma$'s is suppressed either by the small electron Yukawa coupling or by
a loop factor. For $m_S$ below 100 MeV, the decay lifetime $\tau_S$ is given by,
\be
c\tau_S \simeq 1\,{\rm cm} \times \fr{100~\rm MeV}{(\kappa')^2 m_S}.
\ee
Notice that the production of light $S$ states could still be efficient because of 
the rather sizable coupling to nucleons,
\be
{\cal L}_{SN} \simeq   \fr{300~{\rm MeV}\kappa'}{v}\times S\bar NN,
\label{SNN}
\ee
so that the coupling constants participating in the production (300 MeV$/v$) 
and decay ($m_e/v$) can differ by almost three orders of magnitude. 

The new sector coupled to the Higgs portal need not be purely bosonic.
Indeed, one could use secluded fermions $\chi$ attached to $S$ scalars \cite{Bird}, in 
which case the low-energy Lagrangian would also include $\bar \chi \chi O_{\rm SM}^{(h)}$
operators.

\subsubsection*{2.3 Neutrino portal}

The neutrino portal allows for the coupling of a set of singlet fermions $N_j$
to the $LH$ composite fermionic operator of the SM, and results of course in the neutrino Yukawa interaction,
\be
{\cal L}_{\rm int} = y_{ij} L_i H N_j + \cdots,
\ee
where $i,j$ are flavor indices. Together with the allowed Majorana mass terms for the right-handed states $N_j$, these interactions 
can lead to the generation of observable neutrino masses and neutrino oscillation phenomenology, 
and therefore are well motivated.  The flavor structure, which is an important aspect of the neutrino
portal, is of course the primary means via which this portal is probed. Details of neutrino oscillations allow
information to be deduced about the existence of this coupling even though actual detection of 
the right-handed states is clearly out of reach if these states are heavy, as is motivated in the context
of the seesaw mechanism for neutrino mass.

Apart from generating the light neutrino masses, the heavy  states $N_j$  can participate more directly
in SM phenomenology, via the induced mixing angle $\theta_R$. Given that the right-handed states are weakly interacting, 
they can be produced efficiently only in the weak decays of mesons. The typical (naturalness-inspired) mixing angle 
suggested by the see-saw relation is $\theta^2 \sim m_\nu/M_R$. For $M_R\sim {\cal O}$(GeV) this is still a
rather small mixing angle, and detecting it would represent a major challenge. Nonetheless, even 
searches for  $N_j$ states coupled via larger values of $\theta$ are well-motivated phenomenologically, and have been 
discussed in the past on many occasions \cite{neutrino_th}.

\subsubsection*{2.4 Higher-dimensional portals}

Models where  higher-dimensional operators 
mediate the connection to a GeV-scale neutral sector are ubiquitous. 
At dimension 5, significant attention has been devoted to axion-like particles, 
\be
\label{aint}
{\cal L}_{\rm int} = \fr{\partial_\mu a}{f_a}\bar \psi \gamma_\mu\gamma_5 \psi + \cdots,
\ee
where $\psi$ represents a generic SM fermion field and 
the mass of the pseudoscalar  $a$ is protected against radiative corrections 
from (\ref{aint}) by the shift symmetry. A departure from the mass-coupling relation dictated
by the QCD axial anomaly, $m_a^2 \sim m_\pi^2 f_\pi^2/f_a^2$, towards larger values of
$m_a$  allows the avoidance of various prohibitive astrophysical and cosmological constraints. 
This is a model where fixed target experiments could hold a significant 
advantage over colliders,  especially in the kinematic range $m_a < 2 m_\mu$ 
where production is dictated by sizable 
nucleon axial-vector currents, and the decay lifetimes are prolonged by 
chirality suppression, $\Gamma_a \sim m_e^2m_a/f_a^2$.

Moving up in dimension to dimension six, we encounter a multitude 
of operators connecting vector and axial-vector currents of the 
SM with singlet fermions or bosons from a hidden sector. A prominent 
example of such scenarios is the coupling of a light neutralino LSP to 
SM currents. The signal of light neutralino decay, induced either by R-parity 
violation \cite{Dreiner} or by a cascade to even lighter super-partners (such as the gravitino) 
can also be studied  with fixed target experiments. We will not pursue this subject further 
in this paper, noting that the calculations will in general parallel 
the pair-production of $S$-particles from the effective Lagrangian (\ref{eff_h}).

It is also important to bear in mind that there is no intrinsic constraint that the hidden
sector be weakly interacting. General arguments concerning a UV completion would then
suggest the presence of nonabelian gauge fields, and thus the natural portal
could again arise at canonical dimension 6. However, the presence of strong interactions
in both visible (i.e. QCD) and hidden sectors would allow for the possibility of dangerously
irrelevant operators that could produce lower dimensional portals in the infrared, i.e.
a `pion portal'. We will not discuss such possibilities in detail, but will 
comment briefly on an example of this type momentarily.  One of the commonly discussed scenarios
of a strongly interacting hidden sector is the `hidden valley' \cite{SZ}, where for example couplings 
could arise via a pure-gauge portal \cite{FP},
\be
\label{hidv}
{\cal L}_{\rm int} = \frac{1}{\La^4} {\rm Tr}(G^2){\rm Tr}(G'^2) + \cdots\,,
\ee
with $G$ the gluon field strength and $G'$ the corresponding field strength in the hidden sector.
If this is the leading dimension contribution, i.e. the portal starts at dimension 
8, then according to (\ref{counting}) colliders would have a clear advantage over 
fixed target experiments in probing these sectors. However, in a somewhat 
more complicated QCD-like secluded sector, there is the 
possibility for a dimension 6 operator, connecting quark currents,
$\Lambda_6^{-2}(\bar q_{L(R)}\gamma_\mu q_{L(R)})(\bar q'_{L(R)}\gamma_\mu q'_{L(R)})$. 
Such an operator would then translate into the interactions of `hidden 
pions' with the axial-vector SM current, and so would be essentially identical to 
(\ref{aint}). In this case, a long-lived `hidden pion' would be a good candidate 
for fixed target searches.

\subsection*{3. Fixed target signatures of hidden sectors}

We begin this section with a discussion of production and detection of hidden sector particles at proton fixed target experiments.  In general there are 
two mechanisms by which hidden sector states $Y$ will be produced: 1) direct production $p + A \rightarrow Y + X$, 
and 2) production of secondary hadrons $p+A \rightarrow H + X$ followed by the decay $H\rightarrow Y + X$. In general 
the number of candidate events can be written as
\begin{equation}
 N_{\rm events} = N_Y \times P_{\rm det},
\label{events}
\end{equation}
where $N_Y$ is the total number of $Y$ particles produced in the experiment and $P_{\rm det}$ is the probability that 
a $Y$ particle intersects and decays within the detector. We discuss each of these in turn.

\bigskip

{\it (i) Direct partonic production}:
For direct production, the number of particles produced is simply the cross section times the integrated luminosity:
\begin{equation}
 N_Y = \sigma \times N_{\rm POT} \left(\frac{N_A \rho L}{A} \right),
\label{numberprod}
\end{equation}
where $N_{\rm POT}$ is the number of protons on target, $N_A$ is Avogadro's number, $\rho$ and $A$ are  the density and mass number of the material comprising the target, and $L$ is the length of the target.
For direct production there are essentially two regimes of interest. When the kinetic energy of the protons in the beam is large, such that $\sqrt{s} \gg {\rm few}$ GeV, then the production cross sections can be computed using perturbative QCD. On the other hand, for low energy beams a more appropriate description of scattering is based on the baryonic and mesonic degrees of freedom. 
For order-of-magnitude estimates of the number of events, 
the subtleties of hadronic model-dependence are inessential. However, precise predictions generally require a parameterization of the 
production cross section for a given process, based ideally on data from dedicated experiments, as is done for neutrino oscillation 
studies, as well as on detailed simulation of the beam-target interaction and subsequent particle propagation.

\bigskip

{\it (ii) Secondary hadronic decays}:
For $Y$ particles arising from the decays of secondary hadrons $H$, we can write
\begin{equation}
 N_Y = N_H \times {\rm Br}_{H \rightarrow  V X}.
\label{numberbranch}
\end{equation}
To obtain the number of hadrons $N_H$ produced at an experiment, one can 
calculate the production cross section times luminosity directly. 
However we will often find it convenient and perhaps 
more reliable to base our estimate of $N_H$ on 
known experimental quantities, such as the measured neutrino flux.

The probability $P_{\rm det}$ that a produced $Y$ particle reaches and decays or scatters in the detector involves a convolution of the 
angular and momentum distribution of the produced $Y$ particle with the decay probability,
\begin{equation}
P_{\rm decay}={\rm exp}\left(-\frac{d_1}{\gamma v \tau}\right )-{\rm exp}\left(-\frac{d_2}{\gamma v \tau}\right ),
\label{pdecay}
\end{equation}
where $d_1$($d_2$) is the distance from the production point to the entry (exit) point of the detector, $\tau$ is the lifetime of $Y$, and $v$ ($\gamma$) is the velocity (boost) of the particle. Clearly, the convolution is necessary because the distances $d_{1,2}$ depend on the outgoing direction of the $Y$ particle (as well as the detector/beamline geometry), and the velocity and boost are determined by the $Y$ particle momentum. The integration is restricted to angles subtending the detector. The angular and momentum distributions can be calculated in principle, but for low energy proton beams this is not always practical due to the complications of the strong interaction. Again in this case, it is useful to use a combination of data and theory to obtain a parameterization of the distributions, and base predictions on this parameterization. 

To obtain a simple order-of-magnitude estimate of the number of candidate events, which we will be content to do for most of the examples in this paper, we can
make the simplifying assumption that the production of $Y$ particles is isotropic in the c.o.m. frame with a characteristic c.o.m. 
momentum $p_{\rm c.o.m.}$ depending on the details of the process. This leads to a simple expression for the probability $P_{\rm det}$:
\begin{eqnarray}
 P_{\rm det} &\sim& \left(\frac{d \Omega_{\rm c.o.m.}}{4\pi}\right) P_{\rm decay} \nonumber  \\
& \simeq & \left(\frac{d \Omega_{\rm c.o.m.}}{4\pi}\right)\left(\frac{\Delta z}{\gamma v \tau} \right),
\label{numberdet}
\end{eqnarray}
where $d\Omega_{\rm c.o.m.} \sim \gamma_0^2 d\Omega_{\rm lab} $  is the solid angle subtended by the detector in the c.o.m. frame, 
with $\gamma_0$ the boost used to transform between frames, and $\Delta z$ is a typical distance traversed in the detector. In the second line above we have made the approximation that the decay length is long compared to the distance from the target to the detector. The approximate formula (\ref{numberdet}) also requires a characteristic velocity and boost of the produced $Y$ particles, which can be  inferred from the kinematics of the decay or production process. 

Taken together, Eqs. (\ref{events}-\ref{numberdet}) are simple formulae useful for obtaining parametric estimates of event rates. In several examples in this paper we have also performed a more accurate numerical calculation of the event rates using a Monte Carlo simulation. More details on these simulations are given below.
We now move on to signatures of specific models, beginning with the \us\ gauge and Higgs$'$ bosons.

\subsubsection*{3.1 Sensitivity to a secluded \us\  sector}

We will focus our attention on the vector portal and perform a detailed study of
the sensitivity to such a sector arising from fixed target proton beams.
Existing and planned neutrino beam facilities provide impressive coverage in the 
relevant (sub-)GeV energy range:
\begin{itemize}
\item $m_{V,h'} < 100$ MeV - LSND
\item $m_{V,h'} < 1 $ GeV - MiniBooNE
\item $m_{V,h'} \ga 1$ GeV - MINOS, (T2K, NOvA, Project X,\ldots)
\end{itemize}

Depending on the mass scale of the hidden sector portal, the dominant production mechanisms for the 
minimal \us\ model are:
\begin{equation}
\begin{array}{lcl}
m_{V,h'} \la m_\pi & \qquad & \pi^0 \rightarrow \gamma V, \gamma Vh'\\
m_{V,h'} \la 400~{\rm MeV} & \qquad & \eta \to \gamma V, \gamma V h'  \\
                    & \qquad & \Delta \to NV \\
m_{V,h'} \la m_\rh & \qquad & \rh^0,\om,\ph  \to  V h',  V \pi^0(\eta) \\
m_{V,h'} \ga 1~{\rm GeV} & \qquad & q+\bar{q} \to V, Vh',\ldots \\
                                  & \qquad & q+g \to Vh', qV, \ldots
\end{array}
\end{equation}
The final line, relevant for a \us\ sector which lies above the hadronic scale, corresponds to direct production
from partonic interactions. In what follows, we will consider the proton beam experiments which provide the best 
sensitivity to secluded vectors and Higgs$'$ bosons. 

\bigskip
\noindent {\it Sensitivity to $V$}:
\bigskip

If $m_V< m_\pi$, the production of $V$ particles in the decays of 
$\pi^0$ is the dominant production mechanism. Indeed, the cross section for
producing $\pi^0$ in proton-nucleus collisions is large, and 
the subsequent decay of $\pi^0$ is electromagnetic, so the production 
of a $V\gamma$ final state, although suppressed by $\kappa^2$ ,
does not involve the small electromagnetic coupling $\alpha$. 
For the secluded U(1)$_S$ model,  kinetic mixing with the 
photon leads to a decay $\pi^0 \rightarrow \gamma V$ with branching ratio,
\begin{equation}
\label{pi0Vg}
 {\rm Br}_{\pi^0 \rightarrow \gamma V} \simeq 2 \kappa^2 \left(1 - \frac{m_V^2}{m_\pi^2} \right)^3,
\end{equation}
with a very similar expression for $\eta$ decay. Another generic advantage of having 
$\pi^0$ production is the abundance of photons and electron-positron pairs created as a result. As these photons and leptons propagate, they may in turn interact with the target material and if possible produce U(1)$_S$ states.
Therefore, the intense proton beam has the attractive feature
that it is also provides a secondary 
lepton/photon beam. 

Neutral pions $\pi^0$ are produced in primary proton-target interactions, in numbers comparable to those of $\pi^+$. In a given fixed target
experiment, if the vector resulting from $\pi^0$ is long-lived, with a decay length on the order of many meters, it may reach the detector and decay.
Experiments with high luminosity, sensitive to the dilepton decay products, can then be used to probe the \us\ parameter space. The highest integrated luminosity ($10^{23}$ POT) was achieved at the Liquid Scintillator Neutrino Detector (LSND) experiment \cite{lsndev}, which ran from 1993-1998 at the Los Alamos Neutron Science Center (LANSCE). Thus, as we will discuss, LSND puts by far the tightest constraints on the low-mass regions of the \us\ parameter space.

According to Eq. (\ref{numberbranch}), we require an estimate of the sample size of neutral pions, $N_{\pi^0}$, produced in the primary proton-target collisions at LSND. Unlike the charged $\pi^+$ production and reaction 
cross sections, for which dedicated measurements have been made in order to gain a precise understanding of neutrino 
sources, direct experimental information on neutral $\pi^0$ production in proton-nucleus collisions is very scarce, in part because the dominant electromagnetic decay products generally do not escape the target. 
Therefore, instead of calculating the $\pi^0$ production directly, we will normalize the number of $\pi^0$ produced  to the number of $\pi^+$, which is directly related to the measured neutrino flux. Since the overwhelming majority of neutrinos arise from decays-at-rest, we can easily estimate the number of $\pi^+$ produced in the lifetime of the LSND experiment:  
\be
N_{\pi^+} = \frac{\Phi_\nu \times A_{\rm det}}{ (d\Omega_{\rm lab}/4\pi)_\nu }\sim 10^{21},
\label{pi+}
\ee 
where $\Phi_\nu = 1.3 \times 10^{14}$ $\nu$ cm$^{-2}$ is the measured neutrino flux at the detector
integrated over the time of operation,  
$ A_{\rm det} \approx 2.5\times 10^{5} $ cm$^{2}$, 
and $(d\Omega_{\rm  lab}/4\pi)_\nu \approx 4 \times 10^{-2}$ is the 
fractional solid angle subtended by the detector relative to the target. 
This is by far the largest sample of charged pions produced in any 
existing experiment.

The total number of neutral pions should be within an ${\cal O}(1)$ factor of the number of $\pi^+$ produced at LSND, which 
can be inferred by comparing the production of  $\pi^0$ and $\pi^+$ in proton-proton and proton-neutron collisions 
(see e.g. \cite{piexp}), or by using simple counting arguments related to the number of production channels \cite{picount}.

From Eqs. (\ref{events},\ref{numberbranch},\ref{numberdet}), and using the 
decay width of $V$ to electron-positron pairs, 
\begin{equation}
\Gamma_{ V \rightarrow \overline{l}l }=\frac{1}{3} \alpha \kappa^2 m_V \sqrt{1-\frac{4 m_l^2}{m_V^2}}
\left(1+\frac{2 m_l^2}{m_V^2}\right),
\end{equation}
we can give a parametric estimate of the number of decay events in the 
regime where the travel distance exceeds the distance to the detector:
\begin{eqnarray}
N_{events} & \sim & \Phi_\nu V_{\rm det}\left( \frac{\alpha \kappa^4 m_V^2}{m_\pi} \right)\nonumber \\
& \sim & 10^5 \times \left(\frac{\kappa}{10^{-6}} \right)^4 \left( \frac{m_V}{\rm 10~MeV}\right)^2,
\label{lsndvap}
\end{eqnarray}
where $V_{\rm det} \sim 2 \times 10^8 {\rm cm}^3$ is the volume of the LSND detector. 
One observes that the number of events within the 
LSND detector can potentially be very large. 

We have also performed a more accurate numerical calculation of the event rate. For this, we have made the reasonable assumption that the energy and angular distribution, $dP/(dT_\pi d\Omega)$, is qualitatively similar to that of the charged pions and have used a parameterization inspired by Ref.~\cite{lsndpion}. 
Once the distribution of the outgoing pions is specified, it is straightforward to numerically calculate the 
probability, $P_{\rm det}$, that a \us\ vector  passes through and decays in the detector. To this end, we have written a simple Monte Carlo code that produces a neutral pion, decays the pion via $\pi^0 \rightarrow \gamma V$, determines if the vector passes through the detector, and if so, calculates the probability of decaying within the 
detector according to Eq. (\ref{pdecay}). A large sample of events is generated, and the probability $P_{\rm det}$ is calculated by properly weighting and summing over events in which \us\ vectors cross the detector.

\begin{figure}
\centering{
\includegraphics[width=0.9\textwidth]{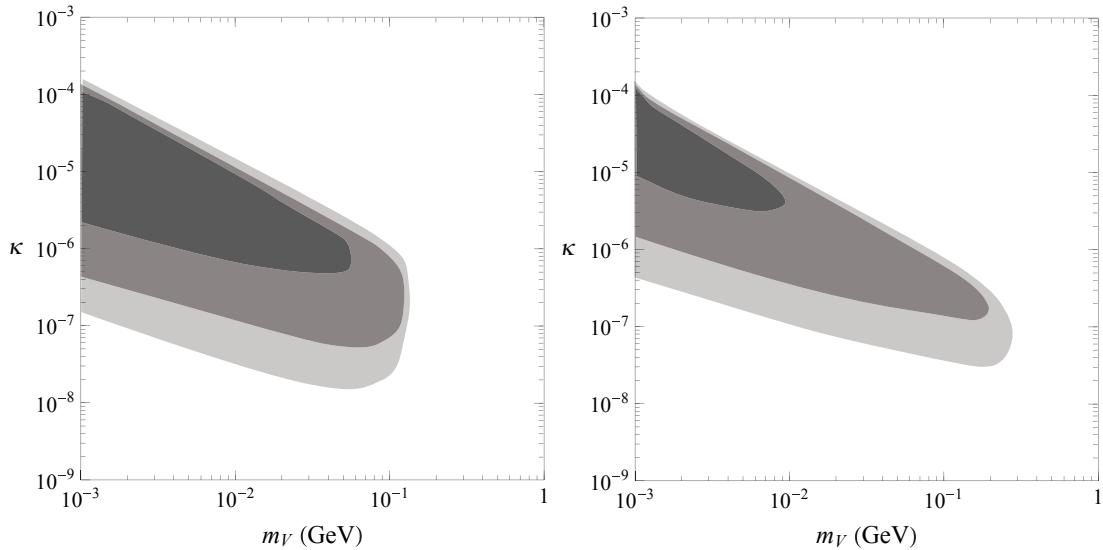}}
\caption{Sensitivity of LSND to decays $V\rightarrow e^+e^-$. The light, medium, and dark shaded regions indicate more 
than 10, 1000, and $10^6$ expected events respectively. The left panel shows events due to vectors arising  from $\pi^0\rightarrow \gamma V$ decays, while
the right panel shows  events arising from $\Delta(1232)\rightarrow N V$.}
\label{fig-lsndvec}
\end{figure}

The total number of events in which \us\ vectors decay within the detector can then be written as
 \begin{equation}
 N_{V-{\rm decays}} \sim \frac{\Phi_\nu A_{\rm det}}{d\Omega/ 4 \pi }\times {\rm Br}_{\pi^0\rightarrow \gamma V} \times P_{\rm det}.
\end{equation}
In Fig.~\ref{fig-lsndvec} we show a plot of the number of expected dilepton decay events over the $\kappa-m_V$ parameter space,
which can be quite significant. The plot shows the result of the full Monte Carlo simulation, but the approximate estimate in Eq. (\ref{lsndvap})
actually produces results which are in agreement with the simulation at the percent level for long-lived vectors. Given the large size of the expected event 
sample, it appears that much of the sensitivity range in Fig.~\ref{fig-lsndvec} can be translated to an exclusion region since electron events from $V\to e^+ e^-$ decays 
should automatically be included in the recorded LSND dataset, and in numbers inconsistent with the well-measured backgrounds from neutrinos
and other sources. However, it is also worth noting that there are additional kinematic distinctions. In particular, the higher energy fraction of 
resulting pairs would have similarities with the decay-in-flight charged-current electron neutrino events \cite{lsndev},
for which the background is small, of  ${\cal O}(10)$. Thus we see that the LSND experiment 
rules out (or is at least sensitive to)  a wide range of masses and mixing angles, $m_V < m_\pi$ and $\kappa\sim 10^{-8} - 10^{-4}$.

In addition to production through decays of neutral pions, vectors can also be produced in the decays of the 
$\Delta(1232)$ resonance, $\Delta \rightarrow N V$, which has a branching of
\begin{equation}
{\rm Br}_{\Delta \rightarrow N V} \simeq {\rm Br}_{\Delta \rightarrow N \gamma} \times \kappa^2 \left(1-\frac{m_V^2}{(m_\Delta-m_N)^2} \right)^{3/2}, 
\end{equation}
where ${\rm Br}_{\Delta \rightarrow N \gamma} \sim 0.005$.
The total number of $\Delta$ particles should be comparable to the total number of pions, as we may anticipate that a 
significant fraction of pions  originate from the decays of $\Delta$. Again we use Eqs.~(\ref{events},\ref{numberbranch},\ref{numberdet}) to 
estimate the number of candidate events with a vector decaying in the detector at LSND, which is shown in the right panel of 
Fig.~\ref{fig-lsndvec}. This production process extends the sensitivity range to $m_V \sim 300$ MeV.

The MiniBooNE experiment \cite{miniboonecomp} at Fermilab  has also accumulated a significant number of  protons on target, of order $10^{21}$, and involves a more energetic 8.9 GeV proton
beam, but a larger 540m path to the detector. 
We have investigated the event rates of vectors originating from $\pi^0$ and $\eta$ meson decays at MiniBooNE and have found the sensitivity to be considerably weaker than for LSND due to the lower luminosity and 
the longer travel distance to the detector, which requires smaller $\kappa$ for the vectors to have a sufficiently long decay distance. These two features also afflict the other neutrino experiments in comparison to LSND as concerns their sensitivity to $V$ decays. For example,  the MINOS experiment \cite{minos} also has $\sim 10^{21}$ POT, with the near detector positioned at a distance of 965m from the target. In contrast, the Higgs$'$ can have a parametrically longer lifetime, and thus higher energy neutrino beams allow sensitivity to a higher mass range through Higgs$'$ decays, a subject to which we turn now.

\begin{figure}
\centering{
\includegraphics[width=.95\textwidth]{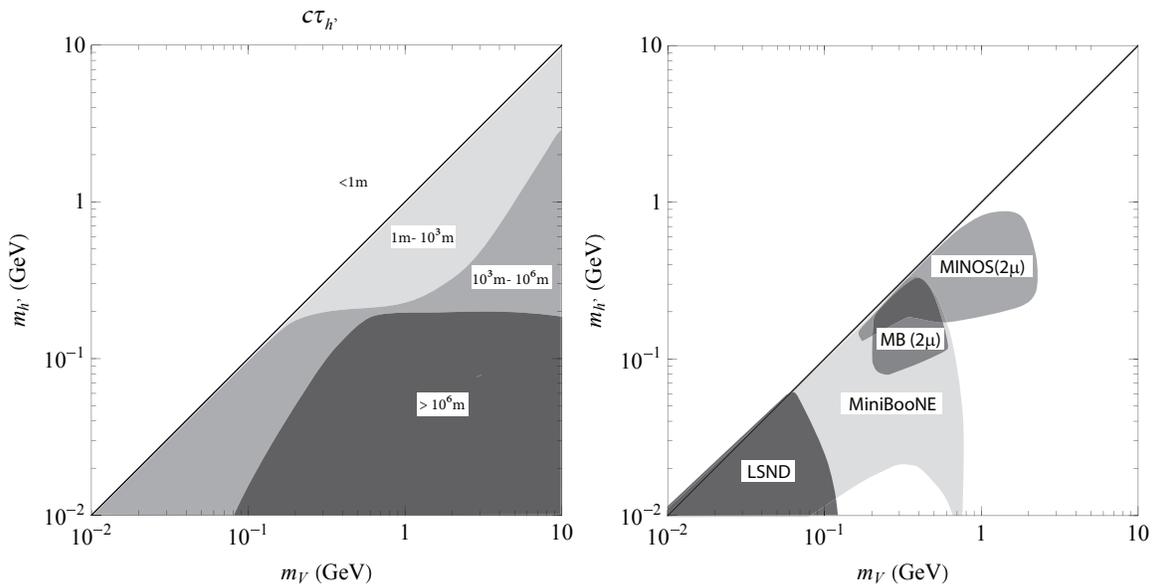}}
\caption{On the left we show  the Higgs$'$ decay length $c \tau$ for $\kappa=10^{-2}$ and $\alpha'=\alpha$. The regions are
$c \tau<1$m (white), 1 - $10^3$m (light), $10^3-10^6$m (medium),
and $> 10^6$m (dark). On the right, we show the sensitivity of LSND ($\pi^0 \to \gamma V h'$ - dark),
MiniBooNE ($\rho,\omega \to h'V$ - light), MiniBooNE ($\rho,\omega \to h'V$,  muon events only - medium), and MINOS (QCD $\to Vh'$,
muon events only  - medium), indicating more than 10 events expected in the detector.}
\label{higgs}
\end{figure}

\bigskip
\noindent{\it Sensitivity to Higgs$'$}:
\bigskip

We now address the production of long-lived  
Higgs$'$ particles at larger values of $\kappa$, where the dominant decay $h' \to l\bar{l}$ occurs at order
$\Gamma_h \sim \kappa^4 \times ({\rm loop~factor})^2$. This requires the kinematic regime
where $m_{h'}< m_V$. Further details of the decay channels which determine the Higgs$'$ lifetime in
different parts of the parameter space are given in \cite{BPR}, and illustrated pictorially in the left panel 
of Fig.~\ref{higgs}. Here we note that specifically 
for  $m_V \gg  m_{h'} \gg 2 m_f$, the lifetime of the $h'$ is
\begin{equation}
\tau_{h'} \sim 6 \times 10^{-9} \, {\rm s } \times \left( \frac{\alpha'}{\alpha}  \right)
 \left( \frac{\kappa}{10^{-2}}  \right)^{-4}
 \left( \frac{m_{h'}}{{\rm GeV}}  \right)^{-1}
\left( \frac{m_V}{2 m_f}  \right)^2.
\end{equation}
[Note that this corrects a typo in Eq.~(14) of \cite{BPR} regarding the dependence on $m_{h'}$.] 
Given the boost in the decay, this allows for a decay length of ${\cal O}(100{\rm m})$ even for $\ka\sim 10^{-2}$. 
 At first sight a value of $\kappa$ 
as large as $10^{-2}$ can be problematic 
due to various independent constraints, such as the size of 1-loop contributions to $g-2$ of the muon and 
the BaBar search for exotic resonances in $\Upsilon(3S)$ decays \cite{BaBar}. However, it is premature 
at this point to declare that $\kappa$ is restricted to lie at or below $10^{-3}$. The measurement 
of the muon anomalous magnetic moment famously disagrees with the SM theoretical prediction at 3$\si$, and 
the existence of a vector with a GeV-scale mass can actually ``help" to resolve the discrepancy \cite{tests}. 
Furthermore, the BaBar analysis has thus far targeted the decays of $\Upsilon(3S)$ to metastable 
pseudoscalar particles, and the analysis makes certain explicit assumptions
that are inconsistent with the \us\ model.\footnote{In fact, as the leptonic widths of these resonances are three orders of magnitude smaller
than the beam energy spread, the production of 
particles from the \us\ sector is not enhanced by any of the $\Upsilon$ resonances at B-factories, and it is the
non-resonant production of $V\gamma$ and $Vh'$ states that dominates,
as {\em e.g.} the $\Upsilon \to \gamma V$ decay is strictly forbidden \cite{BPR,Reece}.}
Therefore, a dedicated search for the \us\ sector  with the full ($\Upsilon(4S)$) dataset is well motivated 
\cite{BPR,slac,Reece}.  Finally, we note that any potential signal involving $h'$ bosons will depend on the secluded gauge coupling $\alpha'$, so that an increase in $\alpha'$ may compensate smaller values of $\kappa'$ and lead to sensitivities similar to those we present below. Here we fix the \us\  coupling to be $\alpha'=\alpha$.

With these issues in mind, we proceed to consider the sensitivity to Higgs$'$ production and decays
for $\ka$ of order $10^{-2}-10^{-3}$. As outlined above, for masses in the GeV range 
the production of Higgs$'$ particles 
from the \us\ sector may proceed in a number of different ways: via the 3-body 
decays of pseudoscalars, $\pi^0 \to \gamma V h'$; via the 
two-body decays of vector resonances, $\rh^0 \to Vh'$, etc; and via direct production
at the parton level, $q+\bar{q} \to Vh'$. 
For completeness, we include the details of the branching fractions and production cross sections for these 
three classes. For three-body decays of $\pi^0$, we have
\begin{equation}
{\rm Br}_{\pi^0 \to \gamma V h'} = \frac{1}{\pi} \int_{(m_V+m_h)^2}^{m_\pi^2} d q^2 
\frac{\sqrt{q^2} \, \Gamma_{V^* \rightarrow V h}(q^2)}{(q^2-m_V^2)^2}\times {\rm Br}_{\pi^0\rightarrow \gamma V^*}(q^2),
\end{equation}
where ${\rm Br}_{\pi^0\rightarrow \gamma V^*}(q^2)$ is given by Eq.(\ref{pi0Vg}) with the replacement $m_V^2\rightarrow q^2$, and 
\begin{equation}
\Gamma_{V^* \rightarrow V h}(q^2) = \frac{1}{12} \alpha' \sqrt{q^2} f\left(\frac{m_V^2}{q^2},\frac{m_{h'}^2}{q^2} \right),
\end{equation}
where we have defined
$f(\hat{x},\hat{y}) =  \sqrt{ \lambda(1,\hat{x},\hat{y}) } 
(\lambda(1,\hat{x},\hat{y}) +12 \hat{x} ) $
in terms of the familiar kinematic function $\lambda(p,q,r)=p^2+q^2+r^2-2pq-2pr-2qr$.
For two-body decays of vector resonances such as $\rh^0$, we have
\be
 {\rm Br}_{\rh\to Vh'} = {\rm Br}_{\rh\to e^+e^-}\ka^2 \left(\frac{\al'}{\al}\right)\left(1 - \frac{m_V^2}{m_\rho^2}\right)^{-2} f\left( \frac{m_V^2}{m_\rho^2}, \frac{m_{h'}^2}{m_\rho^2} \right).
\ee
We also note that the production of vector mesons in the primary proton-target collisions is generally smaller than $\pi^+$ production by a factor of order 100 (see {\it e.g.} \cite{piexp}), which we have accounted for in our estimate of the event rate. 

For more energetic proton beams, such as the 120~GeV NuMI source at MINOS \cite{minos}, direct Higgs$'$ production at the parton level can also  be significant. 
The parton level cross-section for $q\bar{q}\to Vh'$ takes the form,
\be
\hat{\sigma}_{q\bar{q} \to V h'}(\hat{s}) = \frac{Q_f^2\pi \alpha \alpha' \kappa^2 }{9 \hat{s}} 
\left(1 - \frac{m_V^2}{\hat{s}}\right)^{-2} f\left( \frac{m_V^2}{ \hat{s} }, \frac{m_{h'}^2}{\hat{s}} \right),
\ee
where $\hat{s}\simeq s x_1 x_2$ is the center-of-mass energy of the two partons with momentum fractions $x_1$ and $x_2$. The total cross-section for a proton
colliding with a nucleus of atomic number $A$ is $\si(p+A) = Z\si(p+p) + (A-Z)\si(p+n)$, where individual hadronic cross sections require convolution with the parton
distribution functions (pdfs),
\ba
 &&\si(p(P_1) + N(P_2) \to V+h'+X) \\
  && \;\;\;\;\;\;\;\; = \int_0^1 dx_1 \int_0^1 dx_2 \sum_f f_f^{(p)}(x_1) f_{\bar{f}}^{(N)}(x_2) \cdot \si(q_f(x_1P_1) + \bar{q}_f(x_2P_2) \to V + h'),\nonumber
\ea
in which $N$ denotes either a proton or a neutron. We have used the  CTEQ6.6M pdfs \cite{cteq} in the evaluation of the cross sections. The factorization scale
was chosen roughly at the center-of-mass energy for the NuMI beam/target system. We find that the cross section 
for direct production varies from 0.01pb -- 2pb as 
the Higgs$'$ mass drops from 2~GeV down to a few hundred MeV. We should also note that direct production of vectors is also possible via the
resonant process $q\bar{q}\to V$, but given the sensitivity reach in $\ka$ the decay length of the vector is always too short to be relevant for the 
experimental setup at MINOS, where the near detector - approximately a 7m long cylinder with radius 1m - is located 965m from the graphite target. 

Using these production processes and the details of the Higgs$'$ lifetime, we have explored the sensitivity of a number of neutrino experiments to Higgs$'$ decays in the detector. The results are shown in the right panel of Fig.~\ref{higgs} for $\ka=10^{-2}$. These estimates have been obtained using the approximate formulae in Eqs.~(\ref{events}-\ref{numberdet}). We have also checked these results for the case of LSND and NuMI/MINOS 
by running a Monte Carlo code that incorporates the required branchings/production cross section and kinematics, the decay length of the Higgs$'$, and the geometric acceptance of the detectors, obtaining very similar sensitivities to those depicted in Fig.~\ref{higgs}. Given the distance to the (near) detectors, the proton sources at LSND, MiniBooNE and NuMi/MINOS do not have enough luminosity to probe much lower values of $\ka$ via the Higgs$'$ channel. However, we observe an important complementarity with searches for vectors, in that the more energetic proton sources allow this channel to probe masses up to the few-GeV range from dimuon events in the MINOS near-detector.

One final process we will briefly discuss is $h'-V$ conversion in the detector. If the decay length is longer than the distance to the detector, it is possible for the $h'$ to scatter off a nucleus or electron and convert to a vector, which instantly decays to a pair of leptons. We have estimated this signal and there are indeed a large number of events expected at LSND for $\kappa = 10^{-2}$. However, the sensitive kinematic range is limited to lighter vectors and Higgs$'$ bosons, essentially as the Higgs$'$ must have sufficient kinetic energy to upscatter into the heavier vector (recall $m_V > m_{h'}$ in order for the Higgs$'$ to be long-lived). We conclude that LSND is sensitive to Higgs$'$ conversion, but in a region of parameter space already covered by $h'$ decays. Moving to smaller values of $\kappa$, the rate of conversion in the detector is too small to expect a significant number of events.

Thus far we have considered the minimal vector portal scenario, which is justified because $V$ and $h'$ comprise a primary set of states forming a module in any more complex hidden sector, and thus lead to the most generic predictions for experimental sensitivity. However, it is also worth considering which extensions might lead to a sensitivity reach in parts of the parameter space, e.g. $\ka\sim 10^{-4}-10^{-5}$, which fall somewhere
in between the natural range for $V$ and $h'$ decays. One obvious possibility would be a hidden sector with
at least two new gauge bosons $V_1$ and $V_2$, e.g. as part of some nonabelian gauge group,
or simply two distinct \us\ sectors. If the heavier vector state were produced on-shell 
and subsequently decayed $V_2 \to V_1 h'$, with the Higgs$'$ long-lived, this would provide an additional
window into the parts of the parameter that are inaccessible in the minimal model.

\bigskip
\noindent{\it Probing short-lived \us\ particles using a $K_L$ beam}:
\bigskip

Another means of producing long-lived vector states arises via a kaon beam, that
may be obtained in many fixed target experiments. Previously, the 
production of $V$ bosons in the decays of $K^\pm$ was calculated in \cite{tests}.
Here we note that the rare decays of $K_L$ can also be used to constrain the \us\ sector; 
these decays occur well away from the target and may therefore have very low 
backgrounds from other sources. This opens up the  possibility of 
probing $\ka\sim {\cal O}(10^{-4}-10^{-3})$ directly in the decays of $K_L$. 
For example, the branching ratio for $K_L\to V\gamma \to l^+ l^- \gamma$ is given by 
\be
{\rm Br}_{K_L\to V\gamma} \simeq 10^{-3} \times \kappa^2 \times \left(1 - \frac{m_V^2}{m_K^2} \right)^3.
\ee
This rare decay can be probed by searching for resonance structures in $K_L \to \mu^+ \mu^- \gamma$,
which has a SM branching ratio of $3.6\times 10^{-7}$, measured recently by the KTeV collaboration \cite{KTeV1}
with almost $10^4$ events. The search for additional resonant structures among these 
events can also test the vector candidate \cite{tests} interpretation for the ``HyperCP resonance" \cite{HyperCP}.
The KTeV collaboration has also made a precise measurement of the $\pi^0\to e^+ e^-$ decay with 
the branching $(6.44 \pm 0.25 \pm 0.22)\times 10^{-8}$  \cite{KTeV2}, and about 800 events 
coming from the $\pi^0$ decay produced in  $K_L\to 3 \pi^0$. 
According to Eq.~(\ref{pi0Vg}), a relatively light vector with $\kappa^2 \sim O(10^{-7})$
should produce roughly  the same number of events  and therefore can be searched for using the 
same dataset. This would represent  a significant gain in sensitivity over B-factories, 
albeit in a rather limited kinematic region.

\bigskip
\noindent{\it  Constraints from hidden sector WIMP dark matter}:
\bigskip

One of the primary motivations for anticipating a hidden sector is the need to explain dark matter, and while exploration
of the portals is formally independent of other features of hidden sector physics, it is useful to consider the independent 
constraints on the parameter space that may arise when the dark sector contains WIMP dark matter $\ch$. If the dark matter
candidate is charged under  U(1)$_S$, then vector exchange-induced form-factors such as the charge radius can lead to 
significant scattering cross-sections on nuclei, and direct detection limits may constrain $\ka < 10^{-6}$ even for $m_V \sim 1\,$GeV.
However,  even a mild splitting of the charged components significantly ameliorates this constraint \cite{bpr-dd}, and so
much larger kinetic mixing is also perfectly consistent with adding a hidden sector WIMP. A number of other
astrophysical and cosmological constraints have been considered recently in connection with the presence of a 
dark matter candidate charged under a hidden U(1). We note that vectors should decay before BBN, limiting the
lifetime to ${\cal O}$(1s). Constraints on the diffuse extragalactic X-ray background
also impose limits on $m_V/m_\ch$ through upscattering of CMB photons due to the annihilation $\ch\ch\to VV$ integrated over
all halos \cite{xray}. The corresponding impact of energy injection during recombination on CMB anisotropies and polarization
has also been argued to limit $m_V/m_\ch$ in these scenarios to be at or above $10^{-4}$ \cite{cmb}. Thus for a 100~GeV WIMP charged under 
the secluded U(1), it seems clear that from a number of constraints, the vector mass should be above about 10~MeV.
Annihilation of WIMPs accumulated in the solar interior into metastable mediators will in addition provide strong sensitivity 
to lifetimes in the millisecond range for metastable \us\ particles.

\subsubsection*{3.2 Sensitivity to the Higgs  and higher-dimensional portals} 

In this section we estimate potential sensitivity to light scalars and pseudoscalars 
coupled to the SM via the Higgs portal, and via higher-dimensional operators. Since 
this is a subject which has a rather vast scope, we limit the discussion here to the production of 
relatively light ${\cal O}({\rm 100 ~ MeV})$ scalars and pseudoscalars.

The production rate for pseudoscalar particles $a$ with masses comparable to the pion mass, and 
with derivative interactions (\ref{aint}) to nucleons, can be obtained by rescaling the pion
production cross section if one is willing to disregard differences in the isospin dependence. 
In this way, we find the total production rate for ${\cal O}(100)\,$MeV pseudoscalars to be
\be
N_{a} \sim N_{\pi } \times \fr{f_\pi^2}{f_a^2} \simeq 10^{-10}\times N_{\pi } \times \left( 
\fr{\rm 10~ TeV}{f_a}\right)^2,
\label{Na}
\ee
and therefore we can normalize the pseudoscalar flux  through the detector to the 
neutrino flux from decay-at-rest. 
The decay rate of pseudoscalars is dominated by the decays to electrons (and photons at the loop level)
and is given by,
\be
\label{adecay}
\Gamma_a = \fr{m_e^2 m_a}{2\pi f_a^2} ~~~\Longrightarrow~~~ c\tau_a = 5~{\rm m} \times 
\left( \fr{f_a}{\rm 10~ TeV}\right)^2 \left(\fr{\rm 100~ MeV}{m_a}\right).
\ee
With the optimal choice for $c\tau_a$ ($f_a \sim 20 $ TeV), there can be up to a billion 
$a$ decays in the LSND detector! In the regime where the decay length exceeds the distance to the detector, 
the expected size of the signal for a 100 MeV pseudoscalar is 
\be
N_{a~ decay} \sim \Phi_\nu \times \fr{V_{det}}{\beta_a\gamma_a c \tau_a} \sim 10^9\times 
\left( \fr{\rm 10~ TeV}{f_a}  \right)^4,
\label{Nadecay}
\ee
where we took $\beta_a\gamma_a \sim 2$. This estimate shows that even a scale as high as $500 $ TeV 
can be probed via the decay of  axion-like particles. This event rate can also  be translated into constraints on
``pions from hidden valleys", should they have a similar kinematic mass range. More specifically,
the effective decay constant $f_a$ can be identified with $f_{h\pi}/\Lambda_6^{2}$, where 
$1/\Lambda_6^{2}$ is the normalization of the effective operator connecting the axial-vector currents in the 
visible and hidden quark sectors and $f_{h\pi}$ is the ``hidden pion" decay constant. 

Finally, we discuss the sensitivity to scalars $S$ via coupling through the 
Higgs portal. This topic is closely related to hadron physics in the $0^+$ channel, and as such is
shrouded in controversy.  At LSND energies, $S$-particle production
cannot occur from the decay of hadronic resonances. In order to obtain an estimate, we 
use a simple-minded factorization formula for the process $p+N \to p +N +S$, 
together with the coupling constant from Eq.~(\ref{SNN}):
\be
\sigma_{S} \sim \fr{(10^{-3}\kappa')^2}{4\pi^2} \times \sigma_{tot}~~~
\Longrightarrow~~~ N_{S} \sim 10^{-8} \times (\kappa')^2 \times N_\pi. 
\ee
The decays for $m_S \simeq 100$ MeV are given by the same formula as for pseudoscalars,
Eq.~(\ref{adecay}), swapping $f_a^{-1} \to \kappa'/(2v)$. 
The optimal decay range for the scalars, maximizing the LSND signal, is 
$(\kappa')^2 \sim 10^{-4}$, and therefore the number of 
decay events from scalars coupled to the SM via the Higgs portal can maximally reach 
${\cal O}(10^7)$. Although this is an impressive level of sensitivity, the 
rare decays of $K$ mesons \cite{PRV} can probe $(\kappa')^2 \sim {\rm few}~10^{-7}$ 
which suggests rare decays may be a few orders of magnitude more sensitive than 
fixed target probes. MiniBooNE and other more energetic experiments can access an additional 
channel for producing scalars via $\eta\to \pi^0 S$ decays. This branching is enhanced 
because of the small $\eta$ width, that decays either electromagnetically or via three pions with
a reduced phase space. A detailed analysis of this possibility falls outside the scope of this paper. 

Finally, we comment on the pair production of $S$ particles due to the effective Lagrangian 
coupling (\ref{eff_h}), followed by the decay of the scalars to leptons. The production rate 
in this case is generally very small, suppressed by the fourth power of the 
Higgs mass in the denominator.  However, one of the interesting modes is $K_L\to SS$, which 
proceeds via a top-$W$ loop and benefits from the large top Yukawa coupling. 
For $\lambda \sim O(1)$, the branching of $K_L$ to a pair of scalars can reach $10^{-8}$, and 
lead to up to a thousand leptonic decays of $S$ in the MiniBooNE detector 
if the decay length is optimized by a suitable choice of $\kappa'$.

\subsection*{4. Probing a dark matter beam}

An intriguing extension of the idea to probe light states in hidden sectors via one of the 
renormalizable portals is to consider the possibility that the dark matter candidate may itself be
light enough to be produced in fixed target experiments. This would generate a ``dark matter
beam'' which may register in the neutrino detector via neutral current-like interactions. To demonstrate 
this point we will calculate the expected signal at LSND in the MeV-scale dark matter scenario, 
mediated by relatively light vector particles. Such models were first studied in \cite{BF,Fayet} and 
rephrased in the context of kinetically mixed U(1) models in \cite{PRV}. The original motivation for 
considering dark matter candidates in this mass range come from the seemingly inexplicable 
strong and spatially homogeneous emission of 511 keV photons from the galactic bulge \cite{Integral}. It was
proposed that MeV-scale WIMPs annihilating to electron-positron pairs could in principle source 
this emission \cite{Boehm}. 

One of the viable models posits scalar dark matter \cite{PRV} charged under the secluded 
\us\ having kinetic mixing with photons. One considers a regime distinct from that considered above 
with $m_V > m_\chi$, where $\chi$ stands for the dark matter scalar. 
The fact that the DM particles are scalars ensures that annihilation 
proceeds in the $p$-wave, which is rather important as it allows the use of the same 
 mechanism for the annihilation at freeze-out  and for the galactic annihilation 
 to positrons, which has to be much smaller. The freeze-out requirement translates 
to \cite{PRV}:
\be
\fr{\alpha'\kappa^2}{\alpha} \times \left( \fr{\rm 10~ MeV}{m_V} \right)^4 \times \left ( \fr{m_\chi}{\rm MeV} \right)^2\sim 10^{-6}.
\label{MeVfz}
\ee
From now on, for definiteness we take the WIMP mass $m_\chi =1$ MeV in order to be fully 
consistent with the limits on the shape of the 511 keV line \cite{shape} and with constraints
on the accompanying $\gamma$-emission \cite{Yuksel}. 

An important feature of this construction is the dominance of the 
invisible width of $V$ to a pair of DM particles relative to the visible 
width to leptons,
\be
\fr{\Gamma_{ \rm vis}}{\Gamma_{\rm inv}} = \fr{\alpha\kappa^2}{\alpha'}\ll 1.
\ee
At first sight this should greatly reduce the chances for observing the products of $V$ decays,
as opposed to models where the decays of $V$ are dominated by the branching to SM particles. 
However, this need not be the case in situations where there is some advantage in having
a long-lived (or in this case stable) beam of particles that can scatter in the detector.

\begin{figure}
\centering{
\includegraphics[width=.48\textwidth]{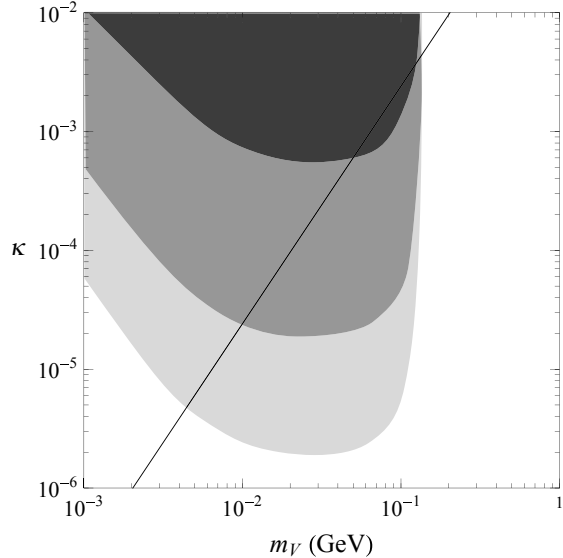}}
\caption{Expected number of neutral current like electron events induced by MeV-scale dark matter scatterings in the LSND detector. We show regions with greater than 10 (light), 1000 (medium), and $10^6$ (dark) expected events. Regions below the black line correspond to a strong coupling regime, with values of $\alpha'> 4\pi$.}
\label{dmbeam}
\end{figure}

In what follows we estimate the signal in the LSND detector from the 
following sequence of events:
\begin{enumerate}
\item $p+p \to X + \pi^0$
\item $\pi^0 \to \gamma V$
\item $V \to 2\chi$ 
\item $\chi + e \to \chi + e$
\end{enumerate}
The first three processes in this sequence occur inside the target and the beam stop, while the 
last process by which the energetic electron is created, occurs within the detector after the dark matter
beam has propagated from the target. The model favors a light vector mediator, which could 
therefore be produced in pion decays. Using the same assumptions 
as in the previous section, and neglecting the phase space factors, we arrive at the 
following estimate for the integrated DM flux through the LSND detector:
\be
\Phi_\chi \sim \kappa^2 \Phi_\nu.
\ee
In order to obtain the number of neutral current-like electron scattering events within 
the detector, we calculate the scattering cross section directly:
\be
\fr{d \sigma_{e\chi\to e\chi}}{dE_f} = \fr{\alpha' \kappa^2}{\alpha} \times 
\fr{8\pi \alpha^2 m_e (1-E_f/E)}{(m_V^2 + 2 m_e E_f)^2},
\ee
where $E_f$ and $E$ are the energies of the recoil electron and incident 
dark matter particles respectively.  

If we conservatively set $m_V ~ \sim 10 $ MeV and $ \alpha' \kappa^2 \sim 10^{-6}$,
then with  $E\sim 100$ MeV we find the total scattering cross section to be
$\sigma_{e\chi\to e\chi} \sim 10^{-34} {\rm cm^2}$. Combining this with the total WIMP flux, the 
number density of electrons in the detector $n_e$,  and the detector volume, 
we arrive at the following estimate for the total number of neutral current-like events:
\be
N_{NC'} = n_e \times V_{det} \times \sigma_{e\chi\to e\chi} \times \Phi_\chi  \ga 10^6.
\label{Nrecoil}
\ee
This result is three orders of magnitude above the total number of energetic electrons 
produced by the neutrino beam and observed in the LSND detector. In Fig.~\ref{dmbeam} we show the number of expected events with electron recoil energies above 20 MeV in the $\kappa$-$m_V$ parameter space. We also show the line below which the dark sector is strongly coupled, $\alpha'> 4\pi$. In fact, constraints on the WIMP self-scattering cross section imply a coupling well below the perturbative bound, which further restricts the parameter space. Therefore, we can safely conclude that this particular model of MeV dark matter is ruled out by the LSND dataset. Even more stringent constraints can be obtained if one considers spectral information on the recoiling electrons. Since the DM 
beam energy is not attenuated at all by the target, the energy of the recoil electrons can be 
larger than the main background produced by neutrinos from stopped pions and muons,
and the number of background decay-in-flight events is estimated to be of ${\cal O}(10)$. 
It is important to investigate whether the estimate (\ref{Nrecoil}) can be 
extended to all MeV-scale dark matter models, which would essentially rule out all analogous 
particle physics scenarios of this type, at least in the form motivated by the galactic 511~keV line. 
A possible modification of the model that may escape the stringent LSND limits would involve
a kinematic constraint so that no $V$ particles could be produced. In this case, more energetic experiments such as 
MiniBooNE should provide valuable constraints on a dark matter beam. 
A detailed analysis of the fixed target constraints on MeV-scale dark matter
models will be presented elsewhere \cite{future}.

\subsection*{5. Conclusions} 

In this paper we have explored the sensitivity of the already expansive fixed target
experimental neutrino physics program to a broader range of portals coupling the SM to
hidden sectors.  We have emphasized that the remaining vector and 
Higgs portals are also well-motivated points on which to focus searches for new physics, 
particularly through their possible connections to the  physics of dark 
matter and other hidden sectors such as supersymmetry breaking. 
As elaborated in sections 3 and 4, existing data from LSND, MiniBooNE and NuMI/MINOS is already very constraining
for a number of scenarios of new physics entering through the vector and Higgs portals, and particular
scenarios for MeV-scale dark matter, motivated by the galactic  511~keV signal, are essentially ruled out
by limits imposed by the LSND dataset. 

In this concluding section we will comment briefly on other fixed target probes and expectations
for future improvements. We have focussed our attention on the neutrino sources with the highest
integrated luminosity in the relevant energy range. However, there are several other experiments
that explore different kinematic regimes. The NuTeV experiment at Fermilab had a very energetic
800 GeV proton beam, and actually observed 3 events consistent with a long-lived state decaying
to two muons, but with asymmetric momenta \cite{nutev}. However, with only $10^{18}$ POT and a 1.4km
distance to the target, this would not be consistent with the minimal \us\ scenarios considered here.
For the lower mass range, i.e. $m_V \la 100$~MeV, electron beam dumps can also provide a competitive source of 
constraints, as recently discussed in the context of the vector portal in \cite{slac2}. In particular, E-137
at SLAC \cite{e137} can provide a reach for low mass vectors down to small values of $\ka$ which
is competitive with the limits discussed here from LSND. In addition, E-141 (SLAC) and E774 (Fermilab) provide constraints
for larger $\ka$ due to the small travel distance to the detector, but only for very low mass vectors \cite{slac2}.
For the heavier mass range, and with longer decay distances, it seems that proton sources hold
a clear advantage. Therefore there is significant complementarity between electron and proton beam
probes of hidden sectors.

While existing facilities already impose significant constraints, the development of long baseline neutrino
experiments with high luminosity proton sources suggests significant improvements in sensitivity
in the near future. The NuMI beam is already in use for the MINOS experiment, and will be upgraded for use with
NOvA which began breaking ground on the site for the far-detector in May 2009. Meanwhile,  beam commissioning
for T2K began in April 2009. A notable feature of T2K in the present context is the possible installation, in addition
to the near detector and beam monitors located 280m from the target,  
of a second near-detector at a distance of a few km that would allow a combined sensitivity to a larger 
mass range for metastable particles up to a few GeV. 
In conclusion, further development of the experimental programme devised to study the physics of neutrino flavor
can at the same time play a significant role in probing the remaining renormalizable portals, and thus will have
a primary role to play in exploring hidden sectors containing at least some states which are light relative to the 
weak scale.

\subsection*{Acknowledgements}

We would like to thank R. Essig, P. Fox, P. Meyers, M. Strassler, L.-T. Wang and I. Yavin for helpful discussions.
M.P. is also grateful to the organizers and participants of the ``Long lived particles at the LHC" workshop,
Seattle, May 3-8, 2009, for providing an inspiring research atmosphere. 
The work of M.P. and A.R. was supported in part by NSERC, Canada, and research at the Perimeter Institute
is supported in part by the Government of Canada through NSERC and by the Province of Ontario through MEDT.

\end{document}